\documentclass[twocolumn,showpacs,preprintnumbers,amsmath,amssymb]{revtex4}
\usepackage{tabularx,graphicx}\begin{document}
\newcommand{\beq}{\begin{equation}}
\newcommand{\eeq}{\end{equation}}
\newcommand{\beqn}{\begin{eqnarray}}
\newcommand{\eeqn}{\end{eqnarray}}
\newcommand{\bmath}{\begin{subequations}}
\newcommand{\emath}{\end{subequations}}
\newcommand{\bra}[1]{\langle #1|}
\newcommand{\ket}[1]{|#1\rangle}

\title{Absence of Josephson coupling between certain superconductors}
\author{J. E. Hirsch }
\address{Department of Physics, University of California, San Diego,
La Jolla, CA 92093-0319}

\begin{abstract} 
It is generally believed that superconductivity can occur in materials irrespective of whether the charge carriers  in the material are electrons or holes.
Here we point out that Josephson tunneling would not occur between superconductors with charge carriers of opposite sign.
Consequently, observation of Josephson tunneling between two superconductors implies that their charge carriers have the same sign.
We propose that this has profound implications for the understanding of superconductivity and in particular is consistent with the theory of hole superconductivity.

\end{abstract}
\maketitle 

\section{introduction}

As clearly expressed by Tinkham \cite{tinkham}, ``the essential universal characteristic of the superconducting state'' is ``the existence of the many-particle condensate wave function $\Psi(\vec{r})$, which has
amplitude and phase and which maintains phase coherence over macroscopic distances.''  Hallmarks of this physics,
predicted by Josephson \cite{joseph} and subsequently amply 
verified experimentally \cite{and,merc,scalreview,japp}, are that a voltage difference $V$ between two superconductors 
connected by a ``weak link'' gives rise to an ac supercurrent of frequency $\omega=2eV/\hbar$,
and that a dc supercurrent can exist in the absence of voltage difference. In this paper we point out that when the two superconductors have charge carriers of opposite sign these phenomena should not occur, and that this fact has fundamental implications for the understanding of superconductivity.

The macroscopic wave function of a superconductor, first introduced by Ginzburg and Landau \cite{gl}, is written as  
\beq
\Psi(\vec{r})=|\Psi|e^{i\phi}
\eeq
where both the amplitude $|\Psi|$ and phase $\phi$ are real functions of position. A macroscopic number of electrons occupies this single quantum
state, all sharing a single quantum phase $\phi$. 
The essential achievement of Josephson was the realization that the   phase degree of freedom  $\phi$ can be experimentally probed through
the ac and dc Josephson effects. 
When two superconductors are connected by a weak link the energy is lowest when their phases lock, i.e. are the same for both superconductors, as shown schematically in Figure 1.
The energy as function of the phase difference is given by
\beq
E=-E_j   cos\Delta \phi
\eeq
with $\Delta \phi=\phi_1-\phi_2$. A dc current through the weak link is associated with a given phase difference between the two 
superconductors according to the relation
\beq
I=I_c sin \Delta \phi
\eeq
and a voltage difference between the superconductors $V$ gives rise to a time-dependent phase difference
\beq
\Delta \phi=\frac{2eV}{\hbar} t
\eeq
and as a consequence an ac current
\beq
I(t)=I_c sin(\frac{2eV}{\hbar}t)   .
\eeq
In the presence of a magnetic field, phase differences are replaced by gauge-invariant phase differences \cite{tinkham}
and a variety of interference phenomena result \cite{and,merc,scalreview,japp}.

 \begin{figure}
 \resizebox{8.5cm}{!}{\includegraphics[width=6cm]{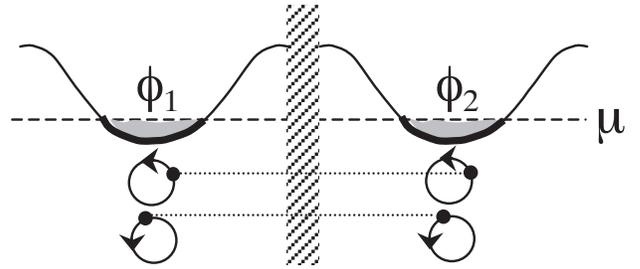}}
 \caption {A tunnel junction or more generally a weak link between two superconductors.
 The thick lines indicate the single particle
 states contributing to the superfluids. On the left  the phase is $\phi_1$, on the right it is $\phi_2$. 
 The phases are locked and evolve with time according to $\phi_i=-2\mu t/\hbar$. When no current flows, $\phi_1=\phi_2$, as shown
 schematically by the rotating dots on the circles.    }
 \label{figure1}
 \end{figure}

The time dependence of the phase of a superconductor follows from the fact that the macroscopic wave function obeys a 
Schr\"odinger equation \cite{gorkov1,anderson}
\beq
i \hbar \frac{\partial \Psi}{\partial t}=2\mu\Psi
\eeq
where $\mu$ is the chemical potential, or energy per particle, hence its time dependence is
\beq
\Psi \sim e^{-2i\mu t /\hbar}  .
\eeq
The 
quantity $|\Psi|^2$ gives the density of superconducting charge carriers which are understood to be Cooper pairs. We denote by
$n_p$ the density of Cooper pairs and $n_s$ the density of individual carriers, hence
\beq
|\Psi|=n_p^{1/2}=(\frac{n_s}{2})^{1/2} .
\eeq
We denote the charge of the Cooper pair and of each member of the pair by $q_p$ and $q_s$ respectively, and their masses by 
$m_p$ and $m_s$. Hence $q_p=2q_s$, $m_p=2m_s$. 
According to the rules of quantum mechanics the 
supercurrent density is given by
\beq
\vec{J}=\frac{q_p n_p}{m_p}(\hbar\vec{\nabla}\phi-\frac{q_p}{c}\vec{A})
\eeq
with $\vec{A}$ the magnetic vector potential. On taking the curl on both sides of Eq. (9),
\beq
\vec{\nabla}\times \vec{J}=-\frac{n_pq_p^2}{m_pc}\vec{B}
\eeq
with $\vec{B}$ the magnetic field, and using  Maxwell's equation $\vec{\nabla}\times \vec{B}=(4\pi)/c)\vec{J}$ yields 
\bmath
\beq
\vec{\nabla}^2\vec{B}=\frac{1}{\lambda_L^2}\vec{B}
\eeq
\beq
\frac{1}{\lambda_L^2}=\frac{4\pi n_pq_p^2}{m_pc^2}=\frac{4\pi n_s q_s^2}{m_s c^2}
\eeq
\emath
describing the fact that the magnetic field cannot penetrate the superconductor beyond a London penetration depth $\lambda_L$. 
From Eqs. (11b) and (8),
\beq
|\Psi| =n_p^{1/2}=(\frac{n_s}{2})^{1/2} \propto \frac{1}{\lambda_L} .
\eeq
As the number of supercarriers decreases the amplitude of the macroscopic wave function decreases and the London
penetration depth increases.

\section{the puzzle}

Let us assume for simplicity a material with a single energy band. When the band is less than half-full it is customary to describe the transport of electricity as being done by electrons, with negative charge, and when it is more than half full by holes, with positive charge. It may be argued that this is only a matter of semantics   \cite{ashcr}, since we can equally well describe the transport for more than half-filled bands with negative electrons, albeit with 
some of them having a negative effective mass. However, we argue that when we are dealing with superconductors the difference is $not$ semantics but has real physical significance.

As discussed above, the amplitude of the macroscopic wavefunction describing the superconductor is the square root of the
number of pairs. Is it electron pairs, or hole pairs, or can we choose either? The answer is, it is $not$ up to us to make the 
choice. In particular, if the
band is almost empty  $n_p$ in Eq. (12) is the number of electron pairs, and if the band is almost full $n_p$ in Eq. (12)  is the number
of hole pairs. This is evident from the fact that in both cases the number of carriers that contribute both to the normal state current
and to the supercurrent is  becoming very small and the London penetration depth   is diverging. 

To make this point clearer, let us consider a simple model of superconductivity, an attractive Hubbard model on a 
simple cubic lattice. 
A straightforward derivation within BCS theory gives for the London penetration depth at zero temperature in direction $\delta$ the relation \cite{kernel}
\beq
\frac{1}{\lambda_L^2}=\frac{8\pi e^2 t}{\hbar^2 c^2 a}\frac{1}{N}\sum_k cos k_\delta[1-\frac{\epsilon_k-\mu}{E_k}]
\eeq
with $N$ the number of lattice sites, $a$ the lattice spacing, $e$ the electron charge,
$t$ the hopping amplitude, $\epsilon_k$ the band energy and 
$E_k$ the quasiparticle excitation energy. From the dispersion relation
\beq
\epsilon_k=-t(cos(k_xa)+cos(k_ya)+cos(k_za))
\eeq
it follows that 
\beq
t=\frac{\hbar^2}{2m a^2}
\eeq
where $m$ is the effective mass for electrons near the bottom of the band when the band is almost empty, or the effective mass
for holes near the top of the band when the band is almost full. For an almost empty band
\bmath
\beq
\sum_k cos k_\delta[1-\frac{\epsilon_k-\mu}{E_k}]\sim \sum_{\epsilon_k<\mu} 1=Na^3 n_e
\eeq
with $n_e$ the density of electrons in the band, and for an almost full band
\beq
\sum_k cos k_\delta[1-\frac{\epsilon_k-\mu}{E_k}]\sim \sum_{\epsilon_k>\mu} 1=Na^3n_h
\eeq
\emath
with $n_h$ the density of holes in the band, 
and from Eqs. (13), (15) and (16)  it follows that  
\bmath
\beq
\frac{1}{\lambda_L^2}=\frac{4\pi n_e e^2}{m c^2}
\eeq
and
\beq
\frac{1}{\lambda_L^2}=\frac{4\pi n_h e^2}{m c^2}
\eeq
\emath
respectively, in agreement with Eq. (11b).

Therefore, {\it the macroscopic wavefunction $\Psi(\vec{r})$ describing the condensate
for the case of an almost full band describes the holes, rather than the electrons in the band.}
We would like to stress that this is by no means obvious. For example, in the early days of superconductivity research it was
widely assumed that when a system goes superconducting the electrons become `free' of interactions with the ionic lattice \cite{early,early2}.
Taken literally this would imply that all the electrons in the band contribute to the superfluid density, rather than just the
`missing electrons' when a band is almost full. That it is the latter is established both by experiments that measure the
superfluid weight or the London penetration depth, and by model calculations within
BCS theory such as the one described above.

As a consequence of this, the amplitude of $\Psi$ for an almost full band gives the density of holes,  {\it the phase of $\Psi$ is the phase of hole carriers rather than that of electron carriers}, and
 the time dependence of the macroscopic wave function for an almost full band is
\beq
\Psi \sim e^{+2i\mu t/\hbar}
\eeq
instead of Eq. (7), which obeys, analogously to Eq. (6), the equation
\beq
i \hbar \frac{\partial \Psi}{\partial t}=2(-\mu)\Psi
\eeq
where $(-\mu)$ is the hole chemical potential, or energy per hole.
 While this is irrelevant for the case of an isolated superconductor, it becomes highly relevant for the case of
a Josephson junction where the superconductor on one side has electron carriers and the one on the other
side has hole carriers.

 \begin{figure}
 \resizebox{8.5cm}{!}{\includegraphics[width=6cm]{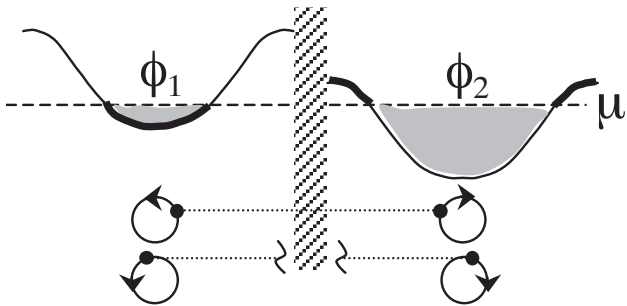}}
 \caption {A  weak link between two superconductors where the carriers are electrons on the left side (side 1) and holes on the right side (side 2).
 The time evolution of the phases is $\phi_1=-2\mu t/\hbar$, $\phi_2=2\mu t/\hbar$. As a consequence the phase difference changes rapidly
 with time, $\phi_2-\phi_1=4\mu t/\hbar$ in the absence of applied voltage, and phase locking would not occur. 
 }
 \label{figure1}
 \end{figure}

Consider then a Josephson junction where the superconductor on the left side (1)
has an almost empty band and the one on the right side (2) has an almost full band, as shown in Fig. 2. The
phases of the macroscopic wave functions evolve in time in opposite directions. The energy  Eq. (2) would now predict
\beq
E=-E_j  cos(4\mu t/\hbar)
\eeq
i.e. an energy that varies rapidly with time in the absence of an applied voltage, and 
as a consequence the phases cannot lock. In this situation we expect
neither the dc nor the ac Josephson effect to take place, and no interference phenomena with
magnetic fields would be seen.

Let us consider an alternative argument to make the same point. A phenomenological derivation of the Josephson
effects starts from the equations \cite{feyn,merc}
\bmath
\beq
i \hbar \frac{\partial \Psi_1}{\partial t}=2\mu_1\Psi_1+\lambda \Psi_2
\eeq
\beq
i \hbar \frac{\partial \Psi_2}{\partial t}=2\mu_2\Psi_2+\lambda \Psi_1
\eeq
\emath
describing a weak  linear coupling between the wave functions on both sides. Assuming the form Eq. (1) for both
$\Psi_1$ and $\Psi_2$ with $|\Psi_i|=n_{pi}^{1/2}$, Eqs. (21) yield (assuming equal superfluid densities on both sides)
\bmath
\beq
\frac{\partial n_{p1}}{\partial t}=-\frac{\partial n_{p2}}{\partial t}=\frac{2\lambda}{\hbar}n_{pi}
sin(\phi_2-\phi_1)
\eeq
\beq
\frac{\partial}{\partial t}  (\phi_2-\phi_1)=   \frac{2(\mu_1-\mu_2)}{\hbar}
\eeq
\emath
which correctly describe the Josephson effects for the situation shown in Fig. 1. However, for the situation shown in Fig. 2 the correct relation instead of Eq. (22a) is 
\beq
\frac{\partial n_{p1}}{\partial t}=+\frac{\partial n_{p2}}{\partial t}
\eeq
because when electrons are transferred from left to right 
the superfluid density decreases on both sides, and conversely when electrons are transferred from right
to left the superfluid density increases on both sides.
To obtain this relation we would have to assume, again with wavefunctions of the form Eq. (1),  the phenomenological equations
\bmath
\beq
i \hbar \frac{\partial \Psi_1}{\partial t}=2\mu_1\Psi_1+\lambda \Psi_2^*
\eeq
\beq
i \hbar \frac{\partial \Psi_2}{\partial t}=2\mu_2\Psi_2+\lambda \Psi_1^*
\eeq
\emath
which  yield 
\bmath
\beq
\frac{\partial n_{p1}}{\partial t}=+\frac{\partial n_{p2}}{\partial t}=-\frac{2\lambda}{\hbar}n_{pi}
sin(\phi_1+\phi_2)
\eeq
\beq
\frac{\partial}{\partial t}  (\phi_1+\phi_2)=  - \frac{2(\mu_1+\mu_2)}{\hbar}
-\frac{2\lambda}{\hbar} cos( \phi_1+\phi_2)
\eeq
\emath
which {\it do not} describe the Josephson effects. In particular, a dc current only results if $\mu_1=-\mu_2$ and
$\phi_1+\phi_2=\pi/2$, and for a non-zero voltage difference between both sides, $\mu_1+\mu_2=2eV$, the
time-dependent current does not have frequency $2eV/\hbar$.

  \begin{figure}
 \resizebox{7.0cm}{!}{\includegraphics[width=6cm]{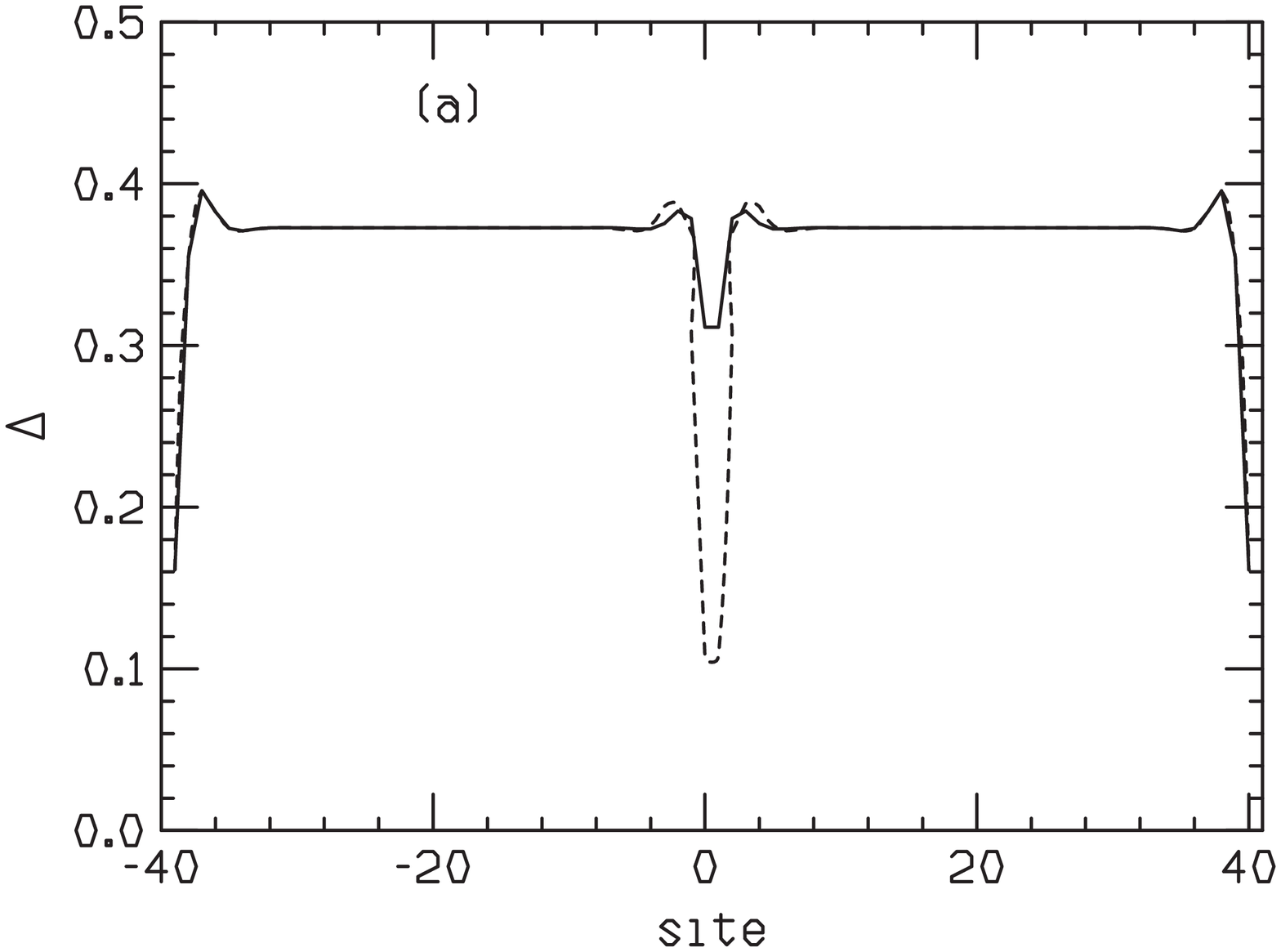}}
  \resizebox{7.0cm}{!}{\includegraphics[width=6cm]{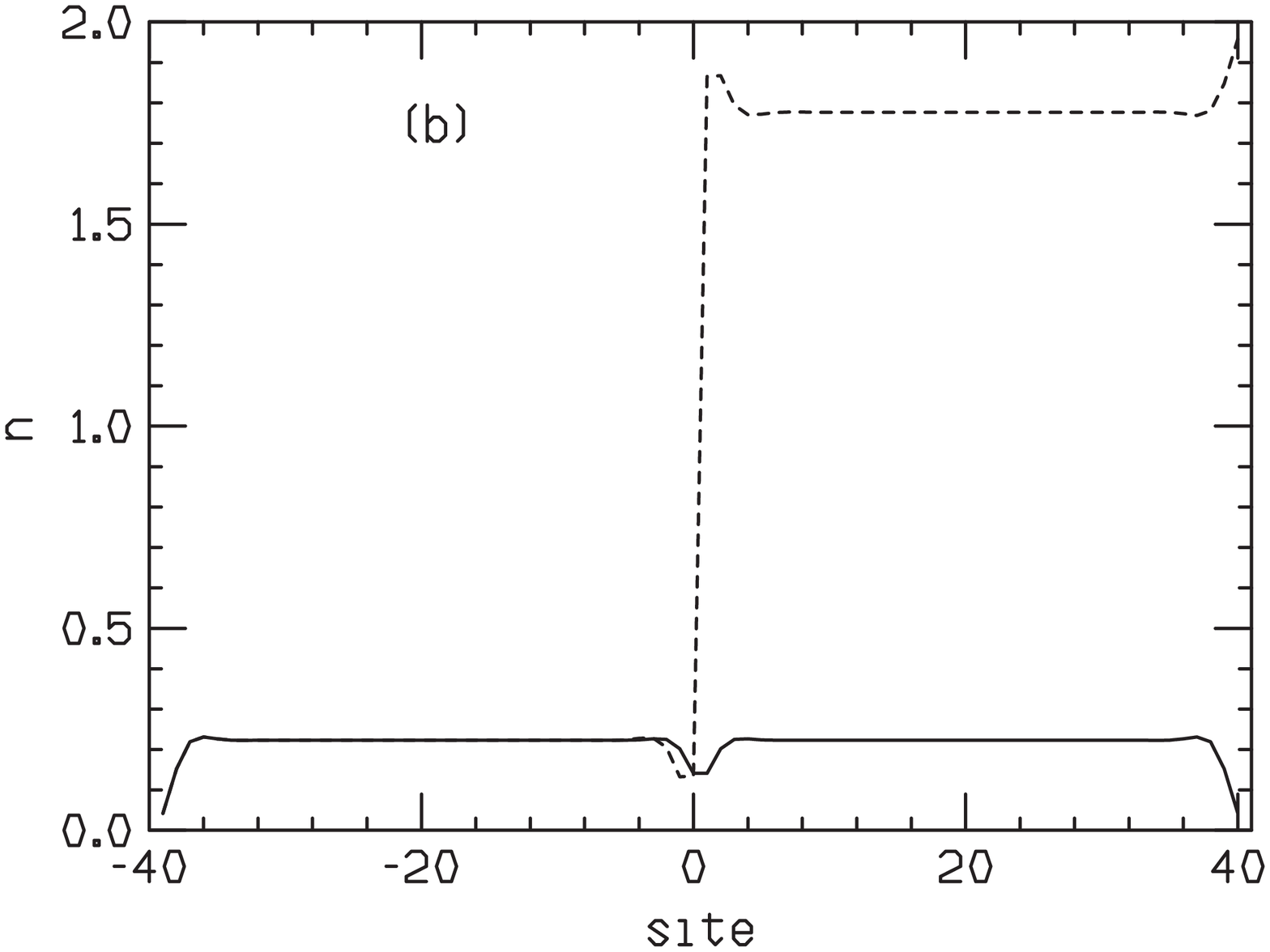}}
 \caption {Bogoliubov-de-Gennes solution for an attractive Hubbard model on a one-dimensional chain
 of 80 sites. Parameters in the model are: on-site interaction $U=-2$ at every site, 
 hopping amplitude $t=1$ between neighboring sites except between sites $0$ and $1$ where
 the hopping amplitude is $t'=0.8$, representing a `weak link', and chemical potential $\mu=0$. 
 (a) shows the gap and (b) the site occupations.
 For the full lines the site energies at all sites
 are $\epsilon=1.9$, for the dashed lines the site energies are $\epsilon_1=1.9$  for
 sites $i\leq 0$, $\epsilon_2=-1.9$ for $i\geq 1$. For the full lines the average site occupation is 
 $0.215$ electrons per site, for the dashed lines the average site occupation is 
 $0.212$ electrons per site for sites $i\leq 0$ and $1.788$  electrons, or $.212$ holes, per site for sites  $i\geq 1$.
 The dashed line is nearly  indistinguishable from the full line   for $i\leq 0$ in (b).}
 \label{figure1}
 \end{figure} 
 
    \begin{figure}
 \resizebox{7.0cm}{!}{\includegraphics[width=6cm]{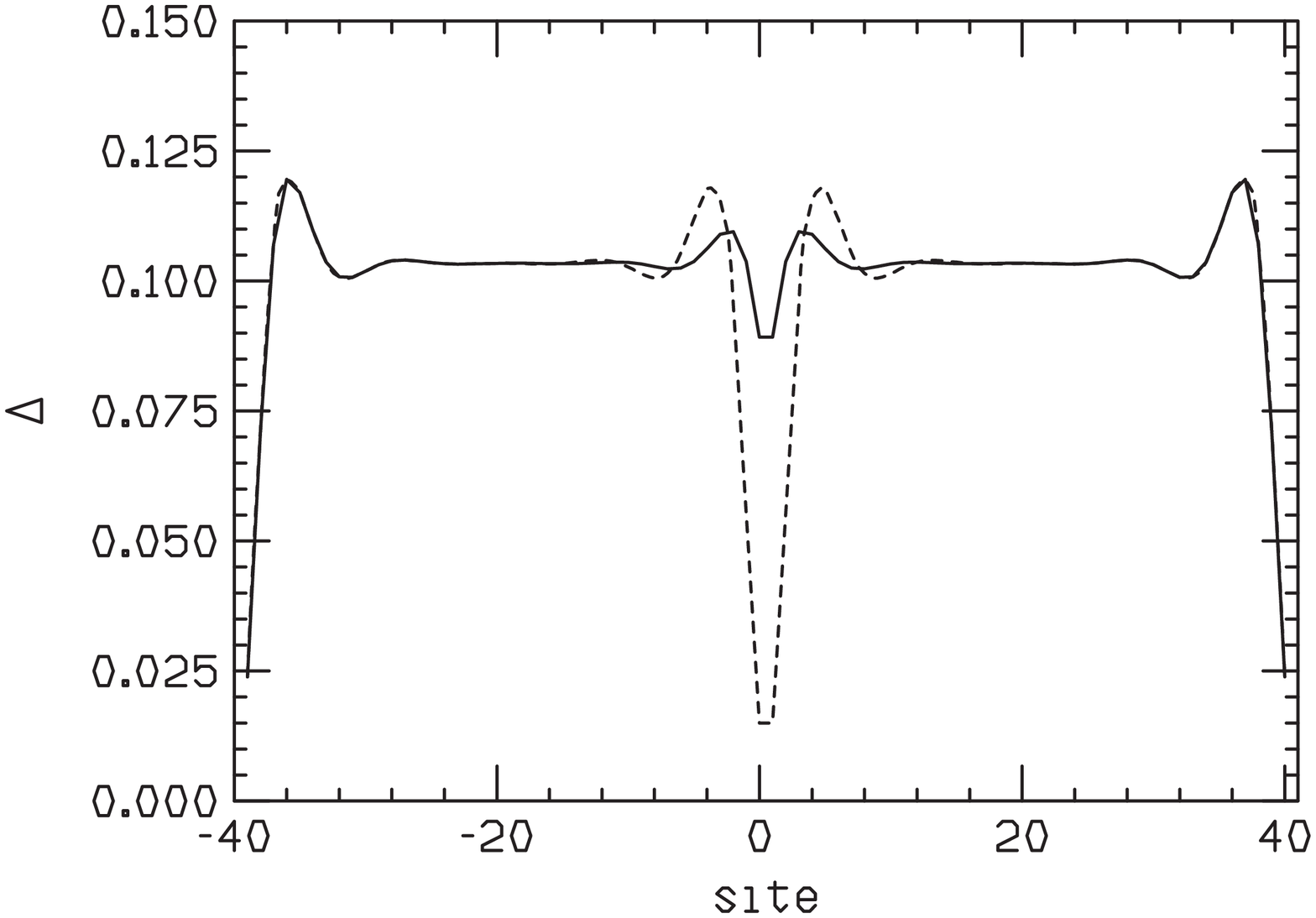}}
  \resizebox{7.0cm}{!}{\includegraphics[width=6cm]{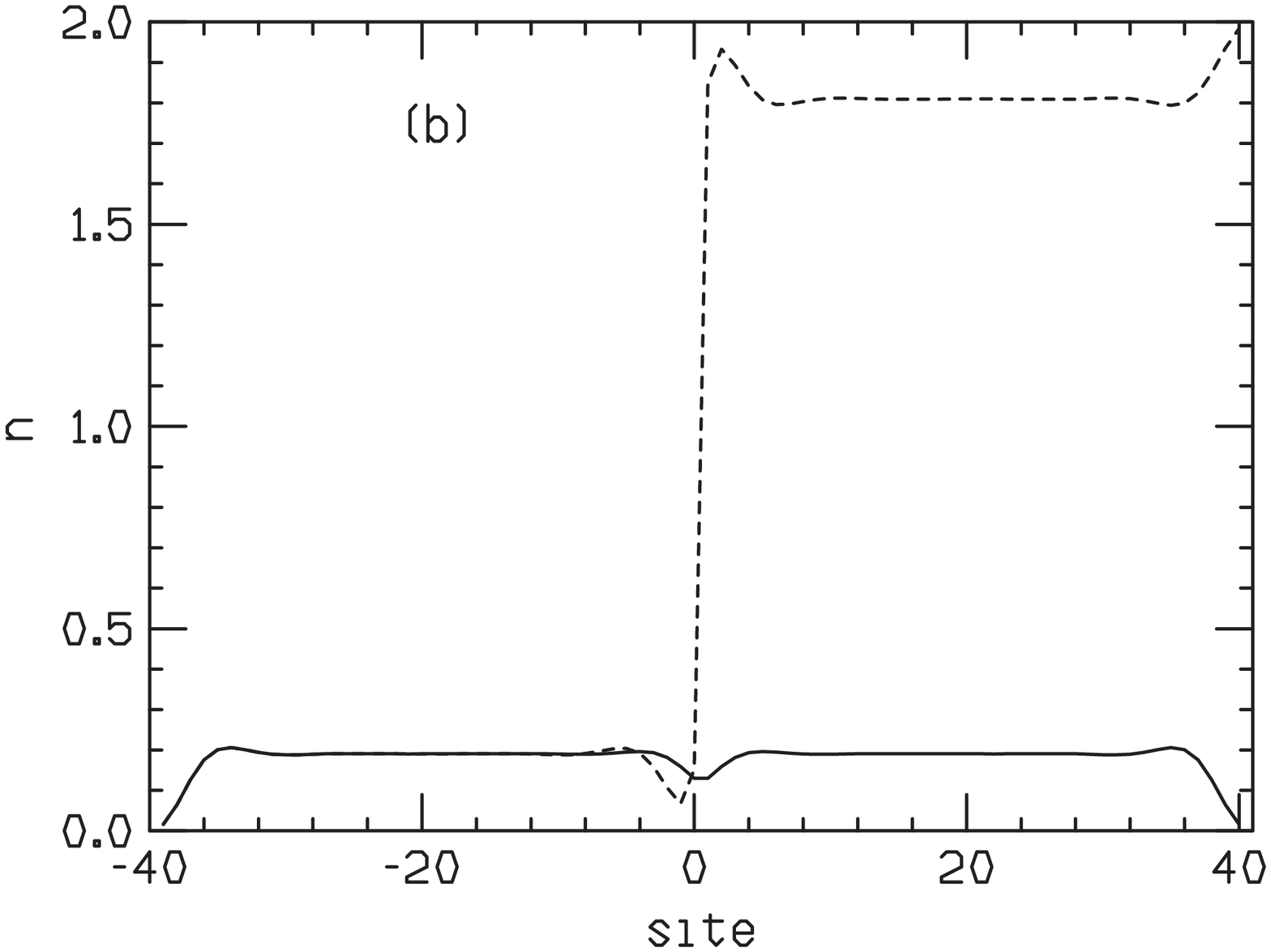}}
 \caption {Same as Fig. 3 except U=-1, t'=0.9.}
 \label{figure1}
 \end{figure}

 We can test these ideas by explicitly solving the Bogoliubov-de-Gennes equations \cite{tinkham} for an attractive
 Hubbard model on a one-dimensional chain, extending from sites $(-N+1)$ to $N$, with a 'weak link'
 between sites 0 and 1 and different site energies on the left and right sides. 
 Figure 3 shows one example, where the full and dashed lines corresponds to the situations shown in
 Figs. 1 and 2. It can be seen that the gap is strongly depressed at the junction when the
 nature of the carriers changes from electron-like on one side to hole-like on the other side. 
 In Fig. 4 we
 show another example with weaker attractive interaction, presumably closer to a real situation, where the
 gap suppression for the case of electrons on one side and holes on the other side is even stronger.
 The gap suppression is  found to be 
 stronger the weaker the attractive interaction is, and in addition we find that it becomes more pronounced when the 
 `weak link' between sites 0 and 1 becomes stronger. The critical
 current in a Josephson junction is proportional to the magnitude of the gaps in the neighborhood of the junction \cite{ab}, 
 so these numerical results indicate that Josephson tunneling is suppressed when 
 the carriers are electron-like on one side and hole-like on the other side of the junction, and particularly when the 
 `weak link' becomes stronger, which is contrary to observations.

 Note also, particularly in Fig. 4, that a charge transfer occurs where electrons migrate
 from the left to the right side of the junction, decreasing the electron occupation on the left and
increasing it  on the right. This is similar to what occurs in a p-n junction between
 semiconductors, where it leads to a depletion layer and rectification at the junction. No rectification effects
 are seen in Josephson junctions to our knowledge \cite{wu}.

In fact however, the usual Josephson effects  are seen to occur with superconductors believed to have
 different types of carriers. For example, Josephson physics has been seen  when one of the
 superconductors is a hole-doped cuprate and the other  one is an electron-doped cuprate \cite{ehcupr}, as well
 as with iron pnictide materials, one with hole carriers and one with electron carriers \cite{ehpnic}. 
More generally,  $if$ the sign of the charge carriers played no particular role in superconductivity as is generally
 assumed, when choosing two superconducting materials at random and connecting them via a weak link, in half of all cases
 we would have one material with electron carriers and the other with hole carriers and
 the Josephson effects would not be seen. That is certainly $not$ consistent with experience  \cite{and,merc,scalreview,japp}.

To make a very clearcut test of this question one could take doped dilute semiconductors that become superconducting.
$SrTiO_3$ becomes superconducting upon doping with n-type carriers \cite{semic1}, $PbTe$ becomes 
superconducting upon doping with holes \cite{semic2}. (Alternatively,  many other
hole-doped group-IV semiconductors exist \cite{semic3}). Unlike the cuprates or pnictides
these  are presumably conventional superconductors where the conventional theory of superconductivity
should apply, with band fillings as shown in Fig. 2. According to the discussion in this paper no
Josephson effects should be seen in weak links or tunnel junctions between doped $SrTiO_3$ and $PbTe$.
If Josephson physics is seen, as we expect it will be, it  will indicate that there is something fundamentally wrong
with the conventional understanding of superconductivity \cite{amiss}.

\section{A possible resolution of the puzzle}

The Josephson effects can be simply understood as arising from the time evolution of the wavefunction of a 
quantum particle in a two-level system \cite{feyn,rogovin,waldram}, corresponding to the two sides of the junction. In the case of Fig. 
1 the ``quantum particle'', represented by the macroscopic wavefunction $\Psi(\vec{r})$, has the same character on both
sides of the junction. In the case of Fig. 2 however the ``quantum particle'' has to change its character from electron to
hole in going from one to the other side, so that  it can no longer be thought of as a single quantum particle in a two-level system.
This explains qualitatively why Josephson couplig cannot occur in this case.

However, the theory of hole superconductivity \cite{holesc} predicts that the situation shown in Fig. 2 can never occur. 
Within this theory,  the sign of the charge carriers plays a fundamental role in
superconductivity \cite{ehasym2}. Superconductivity can only occur if the charge carriers in the normal state
are holes \cite{matmec}, and when the material goes superconducting the character of the
charge carriers changes from hole-like to electron-like, in all superconductors \cite{ehasym,holeelec3}.

Thus, within the theory of hole superconductivity the charge nature of the   carriers in the  normal state
is always the same, and the charge nature of the carriers in the superconducting state
is always the same, independent of the superconducting material. Eqs. (21) apply, and the Josephson
effects should occur between any two superconductors.

For the case of electron-doped cuprates, we have proposed a detailed scenario by which hole carriers get induced
when the system is doped with electrons \cite{edoped}. This scenario is supported by detailed transport measurements 
that indicate that two-band conduction occurs and that superconductivity only occurs when 
hole carriers dominate the normal state transport \cite{greene}. We have proposed that a similar situation occurs in
the pnictide materials \cite{pnic}. Similarly for $SrTiO_3$, hole carriers have to be induced in a band when the system
is doped with electrons in order for it to go superconducting according to the theory of hole superconductivity.

We have pointed out elsewhere \cite{ehasym2,lm} that  the London moment experiments  demonstrate that
superconductors, unlike normal metals, ``know'' the sign of the
charge carriers. In this paper we have pointed
out that the Josephson effects provide further experimental evidence that superconductivity and charge
asymmetry are intimately linked.  Direct experimental evidence of the connection 
of Josephson physics with the physics of the London moment was provided by
Zimmerman and Mercereau \cite{zim}.

If what is proposed here is not the explanation for the puzzle presented in this paper, another explanation should be found.

\acknowledgements
The author is grateful to D. J. Scalapino, A. J. Leggett and P. W. Anderson for comments on the manuscript.

\end{document}